\documentstyle{mn}
\newif\ifAMStwofonts

\newcommand{\be}{\begin{equation}}
\newcommand{\ee}{\end{equation}}
\newcommand{\bea}{\begin{eqnarray}}
\newcommand{\eea}{\end{eqnarray}}
\newcommand{\Bea}{\begin{eqnarray*}}
\newcommand{\Eea}{\end{eqnarray*}}
\newcommand{\pa}{\partial}

\newcommand{\na}{\nabla}

\ifoldfss
  \ifCUPmtlplainloaded \else
    \NewTextAlphabet{textbfit} {cmbxti10} {}
    \NewTextAlphabet{textbfss} {cmssbx10} {}
    \NewMathAlphabet{mathbfit} {cmbxti10} {} 
    \NewMathAlphabet{mathbfss} {cmssbx10} {} 
  \fi
  \ifAMStwofonts
    \ifCUPmtlplainloaded \else
      \NewSymbolFont{upmath} {eurm10}
      \NewSymbolFont{AMSa} {msam10}
      \NewMathSymbol{\upi}     {0}{upmath}{19}
      \NewMathSymbol{\umu}     {0}{upmath}{16}
      \NewMathSymbol{\upartial}{0}{upmath}{40}
      \NewMathSymbol{\leqslant}{3}{AMSa}{36}
      \NewMathSymbol{\geqslant}{3}{AMSa}{3E}

      \let\leq=\leqslant 
       
    \fi
  \fi
\fi 

\ifnfssone
  \newmathalphabet{\mathit}
  \addtoversion{normal}{\mathit}{cmr}{m}{it}
  \addtoversion{bold}{\mathit}{cmr}{bx}{it}
  \newmathalphabet{\mathbfit} 
  \addtoversion{normal}{\mathbfit}{cmr}{bx}{it}
  \addtoversion{bold}{\mathbfit}{cmr}{bx}{it}
  \newmathalphabet{\mathbfss} 
  \addtoversion{normal}{\mathbfss}{cmss}{bx}{n}
  \addtoversion{bold}{\mathbfss}{cmss}{bx}{n}
  \ifAMStwofonts
    \ifCUPmtlplainloaded \else
      \UseAMStwoboldmath
      \makeatletter
      \new@mathgroup\upmath@group
      \define@mathgroup\mv@normal\upmath@group{eur}{m}{n}
      \define@mathgroup\mv@bold\upmath@group{eur}{b}{n}
      \edef\UPM{\hexnumber\upmath@group}
      \new@mathgroup\amsa@group
      \define@mathgroup\mv@normal\amsa@group{msa}{m}{n}
      \define@mathgroup\mv@bold\amsa@group{msa}{m}{n}
      \edef\AMSa{\hexnumber\amsa@group}
      \makeatother
      \mathchardef\upi="0\UPM19
      \mathchardef\umu="0\UPM16
      \mathchardef\upartial="0\UPM40
      \mathchardef\leqslant="3\AMSa36
      \mathchardef\geqslant="3\AMSa3E

      \let\leq=\leqslant 

    \fi
  \fi
\fi 

\ifnfsstwo
  \DeclareMathAlphabet{\mathbfit}{OT1}{cmr}{bx}{it}
  \SetMathAlphabet\mathbfit{bold}{OT1}{cmr}{bx}{it}
  \DeclareMathAlphabet{\mathbfss}{OT1}{cmss}{bx}{n}
  \SetMathAlphabet\mathbfss{bold}{OT1}{cmss}{bx}{n}
  \ifAMStwofonts
    \ifCUPmtlplainloaded \else
      \DeclareSymbolFont{UPM}{U}{eur}{m}{n}
      \SetSymbolFont{UPM}{bold}{U}{eur}{b}{n}
      \DeclareSymbolFont{AMSa}{U}{msa}{m}{n}
      \DeclareMathSymbol{\upi}{0}{UPM}{"19}
      \DeclareMathSymbol{\umu}{0}{UPM}{"16}
      \DeclareMathSymbol{\upartial}{0}{UPM}{"40}
      \DeclareMathSymbol{\leqslant}{3}{AMSa}{"36}
      \DeclareMathSymbol{\geqslant}{3}{AMSa}{"3E}

      \let\leq=\leqslant 

    \fi
  \fi
\fi 

\ifCUPmtlplainloaded \else
  \ifAMStwofonts \else 
    \def\upi{\pi}
    \def\umu{\mu}
    \def\upartial{\partial}
  \fi
\fi
\title[Differential Rotation]
{A Large Eddy Simulation of Turbulent Compressible Convection:
Differential Rotation in the Solar Convection Zone}
\author[Robinson and Chan]
       {Francis J. Robinson
\thanks{Email: marjf@astro.yale.edu}\thanks{Present address: Astronomy Department, Yale University, Box 208101, New Haven, CT 06520-8101.}
 and Kwing  L. Chan\\
       Department of Mathematics, The Hong Kong University of Science and  Technology, Clear Water Bay , Kowloon, Hong Kong}

\pagerange{\pageref{firstpage}--\pageref{lastpage}}

\begin{document}

\maketitle

\label{firstpage}

\begin{abstract}
We present results of two simulations of the convection zone,
obtained by solving the full hydrodynamic equations in a section
of a spherical shell. The first simulation has cylindrical rotation
contours (parallel to the rotation axis) and a strong meridional circulation,
which traverses the entire depth. The second simulation has isorotation
contours about mid-way between cylinders and  cones, and a
weak meridional circulation, concentrated  in the uppermost part of the shell.

We show that the solar differential rotation is directly related 
to a latitudinal entropy gradient, which pervades into the deep
layers of the convection zone. We also  offer an explanation 
of the angular velocity shear found at low latitudes near   
the top. A non-zero correlation between radial and zonal velocity
fluctuations produces a significant Reynolds stress in that  region. This 
constitutes a net transport of angular  momentum inwards, which causes
a slight modification of the overall structure of the differential
rotation near the top. In essence, the {\it thermodynamics controls the
dynamics through the Taylor-Proudman momentum balance}. The Reynolds stresses
only become significant in the surface layers, where they generate 
a weak meridional circulation and an angular velocity `bump'.

\end{abstract}

\section{Introduction: Observations and simulations}
\subsection{Observations}
In the outer 28 $\%$ by radius of the Sun, known as the Solar Convection Zone (SCZ),
most of the vertical energy transport is by convection.
Due to the combination of convection and rotation, the gaseous body exhibits differential
rotation i.e. non-uniform angular velocity. Helioseismology observations (Scherrer et al 1995, Libbrecht 1989)
of the nearly
10 million acoustic p-modes that leak from the interior  into the atmosphere, have provided
through frequency splittings, information on the angular velocity distribution of the solar
interior. The latest results suggest that  the
isorotation surfaces in the solar convection zone are
cone-like (aligned radially), in disagreement with most numerical
simulations, which tend to produce cylindrical contours
parallel to the rotation axis.
Specifically, such observations have revealed that: the angular velocity near the
equator first increases then gently decreases with depth, at mid-latitudes it is almost constant
with depth, while at high latitudes it increases with depth; on the surface of the Sun,
the angular velocity increases from the poles (35 day period) to the equator (25 day period)
and  meridional circulation $v_\theta$,  is much weaker than differential rotation (zonal velocity  $\approx  2{\rm km/s},
v_\theta \approx 25 {\rm m/s}$).

\subsection{Simulations}
\subsubsection{Computational hurdles}
Modeling the SCZ
is a  formidable task.
Observations suggest:
that the SCZ is highly stratified, spanning about 19 pressure scale heights in depth, with
the Mach number (square of the ratio of flow velocity to sound speed)  approaching unity at the top; the Prandtl number (ratio of
the timescales of thermal to viscous diffusion)
is extremely small ($10^{-6}$);
the equation
of state is complex; and finally,
the motion is highly turbulent, revealed by the large number of scales ranging  from
0.1 km to $10^5$ km (depth of SCZ). In the photosphere (near the top of the SCZ) the kinematic viscosity
of hydrogen is about $0.1{\rm cm}^2{\rm s}^{-1}$,
while the root mean square (r.m.s.) velocity in the SCZ is the order of
$100 {\rm m}{\rm s}^{-1}$. These values suggest a  Reynolds number, ${\rm Re}$=velocity $\times$ length /
kinematic viscosity,
of at least $10^{12}$. As the number of degrees of freedom needed to represent a flow is
 proportional to
 ${\rm Re}^{9/4}$, to resolve numerically the scales in all three directions  would
require $10^{27}$ grid points.
 Even with present technology the maximum Reynolds number
 modeled in
direct numerical simulations (DNS) is a few thousand (Brummel, Hurlbert \& Toomre 1996).
 Alternatively, if one assumes most of the energy is contained in the resolved scales, which are much larger than
the viscous and thermal dissipation scales (unresolved scales),
and that the average energy transferred by these smaller scales can be modeled
by entropy diffusion, then a much larger viscosity (smaller ${\rm Re}$) can be used to model the turbulent flow.
This is the principle of
large eddy simulations (LES).

\subsubsection{Numerical simulations  of the SCZ}
Explicit integration of the hydrodynamic equations requires the timestep to be less than 
the time for a  
signal (sound wave) to travel between two grid points. This is known  
the Courant-Friedrichs-Levy (CFL) stability criterion.  
Previous investigations have
solved approximate forms of the fully compressible equations, in which  sound waves do not  exist. 
This circumvents  the use of restrictively small timesteps.
Incompressible convection in a shell was studied by Gilman (1978)  using   the Boussinesq approximation.  This  allows  density 
variation only when coupled with gravity. A better approximation is to exclude only temporal variations in density, known as the 
anelastic approximation. This  
still suppresses sound waves but allows density to vary vertically.
The most recent anelastic simulation in a full shell (global model) on massively parallel architectures
are by Miesch et al (2000). 
Though there were significant improvements on earlier work (Glatzmaier 1987), 
they were still unable to adequately resolve the observed cone-like structure of the angular velocity contours. 
Overall the isorotation contours were cylindrical. 
Furthermore, the angular velocity shear layer near the top of the convection zone 
could not be resolved.  
Near the top, vigorous compressible motions have  
Mach numbers close to one and the anelastic approximation breaks down. 

The fully compressible equations have however been solved in local models (boxes). 
By considering only a tiny section of a spherical shell, 
Brummel et al. (1996)  did a DNS of turbulent compressible convection 
in f-planes.  In essentially the same geometry, Chan (1999)  did a set of  
LES computations, in f-planes.  The LES model emphasized efficient convection  
(convective flux $\gg$ diffusive flux) as opposed to the inefficient convection in the DNS models.
In f-plane simulations 
the angle between the rotation vector and gravity is considered constant throughout the box. 
The  problem with  these local models is that they have periodic horizontal boundaries. In such 
domains the only possible source  of differential rotation
is the Reynolds stress. Horizontal averaging removes latitudinal gradients of thermal quantities  which 
are a possible source of the differential rotation. For example, in the model by Durney (1999), the latitudinal entropy gradient 
(or baroclinicity) is essential in shaping the isorotation contours.     
Additionally, the f-plane models do not allow a realistic  meridional circulation to develop; nor 
can Rossby waves exist as the rotation vector is constant throughout the domain. 
Rossby waves have been  suggested as a possible source of correlation 
between longitudinal and meridional motions (Brummel et al. 1996). 

Our model represents a 
compromise between the anelastic global and compressible local
simulations. 
We solve the full Navier Stokes  equations in  spherical coordinates 
in a domain with a significant latitudinal coverage. In this way   
we are able to resolve the upper shear layers and  simultaneously include the effects of meridional circulation
and latitudinal gradients of thermal quantities. To circumvent the CFL restriction on the time step, 
semi-implicit time integration is employed. We use the Alternating Direction Implicit Method on 
a Staggered Mesh (ADISM) (Chan \& Wolff 1982)   in conjunction with an explicit method. 
Due to computational restraints of three-dimensional calculations, 
we are limited to studying a small section of the entire shell. A full  
simulation on a  $70 \times 70 \times 39$ three-dimensional 
mesh, spans just $60^\circ$ in longitude and latitude. This calculation  
requires about a month to run when parallelized on 4 processors of the ORIGIN 2000.
 
This paper consists of 4 more sections. 
Section 2 describes the overall physical setup, formulates the mathematical model 
and describes the numerical approach.
This is followed by descriptions  of the statistically averaged zonal and
meridional 
flows, and  turbulent nature of the compressible convection.
 The third section attempts to pin down the 
source of the differential rotation and meridional circulation. The final section is 
a conclusion.

\section{Mathematical model}
\subsection{Overall setup}
The computational domain is a 3-dimensional sector of a spherical shell, symmetric about the equatorial
plane. A longitudinal cross-section is shown in Fig. \ref{sdomain}. 
The input energy flux at the bottom (straight arrows), is
transported mostly by convective eddies (curly arrows) up to a height of 95$\%$ of the total
depth. To pad out convective motions, a conduction layer is placed in the upper 5$\%$ of the
shell (shaded region). We consider two models that are identical apart from their longitudinal spans.
The first model, denoted A, spans $30^\circ$ in longitude and $60^\circ$ in latitude, while the second, B, 
covers the same
latitudinal range, but $60^\circ$ in longitude. 
Both cover a total depth of 0.72$R$ to $R$, which
is the approximate depth occupied by the convection zone in the Sun. $R$ is equivalent to 
the radius
of the Sun. As depth is scaled by the total radius, 
the nondimensional radius is between 0.72 and 1.0.
Radial, meridional and  zonal directions are labeled $r$ (increasing outwards), $\theta$ (increasing southwards) 
and $\phi$
(increasing eastwards), respectively.
\begin{figure}
\vspace{5.5cm}
\caption{\label{sdomain} Longitudinal cross-section of the computational shell with rotation axis
${\bf \Omega}$.
The input energy flux (straight arrows) is transported by convection (curly
arrows) and then by radiation (shaded region), to the conducting top.}
\end{figure}

\subsection{The equations}

In the absence of motion ($\pa/\pa t=0$, ${\bf v = 0}$), the equations governing conservation of mass,
momentum and energy in a rotating stratified fluid (Chan 1999),
reduce to the equations of hydrostatic
and thermal equilibrium. A solution of these, in spherical coordinates, is a polytrope 
\bea
 T/T_{\rm t} &=& 1 + Z(R/r -1)/(R/r_{\rm b} -1 ),
\label{poly1}
 \\
\rho/\rho_{\rm t}&=&(T/T_{\rm t})^n,
\label{poly2}
 \\
p/p_{\rm t}&=&(T/T_{\rm t})^{n+1},
 \label{poly3}
 \eea
where $r$ is the radial distance from the base, and $n$ is the polytropic index. The subscripts `t' and `b' denote a quantity
measured at the top and bottom of the shell, respectively. $T, p$ and $\rho$
are the symbols for temperature,
pressure and density. $Z = (T_{\rm b} - T_{\rm t})/T_{\rm t}$ describes the extent of the stratification. 

  Equations ({\ref{poly1}-\ref{poly3}) provide a reference atmosphere from which appropriate dimensionless
units can be formed. Length is scaled by the outer radius of the shell $R $
, and time by
$R/\sqrt{p_{\rm t}/\rho_{\rm t}}$.
 In such units, velocity is scaled by the isothermal sound speed at the top,
$\sqrt{p_{\rm t}/\rho_{\rm t}}$. 
From now on all quantities will be given  in  nondimensional units. 
Combining
 the equation of hydrostatic equilibrium, $dp/dr=-\rho g$ and equation (\ref{poly1}), gives in
nondimensional units, $g_{\rm t} = (n + 1)Z r_{\rm b}/(1 - r_b)$. We will consider $g_{\rm t}$
as an independent
parameter. As the total number of pressure scale
heights is $(n + 1) \ln (1 + Z)$, the size of  $g_{\rm t}$ determines the depth of the layer.

 With such scaling, the governing equations become:

\bea
\pa \rho / \pa t & = &  - \na \cdot  {\bf \rho v },
\label{ns1}
\\
{\bf  \pa \rho v} / \pa t & = &- {\bf\na \cdot \rho  v v}
- \na  p
+ \na \cdot {\bf \Sigma}
\nonumber
\\
& & - \rho  g {\hat {\bf r}}
- 2 \rho {\bf \Omega_o  \times  v},
\label{ns2}
\\
\pa E / \pa t &=&
-\na \cdot \left(\frac{1}{\gamma-1}\rho T {\bf v}+p {\bf v}
 + (\rho {v^2} /2) {\bf v}\right)
\nonumber
\\
& &   -\na \cdot \left( -{\bf v}\cdot {\bf \Sigma}
 + {\bf f} \right)  
-  {\bf {\rho v}} \cdot g {\hat{\bf r}},
\label{ns3}
\\
p&=& \rho T,
\label{ns4}
\eea

where ${\bf \Omega_o}$, ${\bf \Sigma}$ and ${\bf f }$ are defined below.                                                                  
Equations (\ref{ns1}-\ref{ns4}) represent a closed system of 5 dependent variables:
density, radial mass flux,
meridional mass flux, zonal mass flux and total energy density,
denoted by $\rho, \rho v_r, \rho v_\theta, \rho v_\phi$
and $E$, respectively. In a particular geometry (fixed $r, \theta$ and  $\phi$ boundaries), each turbulent
convection simulation is specified by defining  {\it{six}} nondimensional parameters. 
These are the reference rotation rate $\Omega_{\rm o}$, the input energy flux  at the base $f_{\rm b}$,
the turbulent Prandtl number ${\rm Pr} = \nu/\kappa$, the gravitational acceleration at the top $g_{\rm t}$, the polytropic index $n $
and the ratio of the specific heats $\gamma$.

We will now consider individual  terms in equations (\ref{ns1}-\ref{ns4}).
Ignoring the coefficient of bulk viscosity (Becker 1968), the viscous stress tenser 
for a Newtonian fluid
is $\Sigma_{ij}=\mu(\pa v_i/\pa x_j+\pa v_j/\pa x_i)-2\mu/3(\na \cdot {\bf v})\delta_{ij}$.
In DNS, the viscosity is determined
from the nondimensional parameters characterizing the convection simulation (e.g $\rm {Pr}$, the Rayleigh number Ra, $g_{\rm t}, n$  and $\gamma$) . 
However, in LES, 
the
dynamic viscosity $\mu$  is increased so that it represents the effects of Reynolds
stresses on the unresolved or sub-grid scales (SGS) (Smagorinsky 1963),
\be
\mu=\rho(c_\mu\Delta)^2(2 \mbox{\boldmath $\sigma : \sigma$})^{1/2}.
\ee
The colon inside the brackets denotes tensor contraction of the rate of strain tensor 
$\sigma_{ij} = (\na_i v_j+\na_j v_i)/2$.
The SGS eddy coefficient $c_\mu$, is set to 0.2, the value for incompressible
turbulence, and $\Delta^2 = r \Delta r \Delta \theta$ is an estimate of the local mesh size. Numerical experiments with different $\Delta$ and $c_\mu$ 
 are described in  Chan \& Wolff (1982). 
The  present formulation ensures that the grid Reynolds number $\Delta \times v/\nu$  is of order
unity everywhere. 
To handle shocks, $\mu$ is 
multiplied by $1 +2/{c_{\rm s}}^2[(\Delta x \pa_x v_x )^2 + (\Delta y \pa_y v_y )^2]$, where 
$x$  and $y$ denote meridional and zonal directions (e.g. $\Delta x = r\Delta \theta$ and $\Delta y = r {\rm sin} \theta \Delta \phi$) 
and $c_{\rm s}$ is the isothermal sound speed.  
As $\mu$ is dependent on the horizontal divergence, any large horizontal velocity gradients are smoothed out by the increased viscosity.

 The gravitational acceleration at a distance $r$ from the center of the sphere is $g\hat{{\bf r}}$,
 where $g=g_{\rm t}/r^2$, and $\hat{\bf r}$
is a unit vector directed radially outwards. The reference rotation vector is  
${\bf \Omega_{\rm o}} =\Omega_{\rm o} \hat{\bf{\Omega}}$, where
 $\hat{\bf {\Omega}}$  is a unit vector in the direction of the rotation axis,i.e.  
${\hat{\bf \Omega}} \cdot {\hat{\bf r}}={\rm cos} \theta$.
The nondimensional rotation rate  $\Omega_{\rm o}$, is 
equal to the ratio of the solar rotational velocity, to the 
isothermal sound speed at the top of the convection zone. For the sun 
$\Omega_{\rm o}$ is about 0.3, so that 
the  value of 2.91 used in the present computation, is associated with rotational periods of about 
a day. 

The total energy per unit volume $E$, is the sum of the 
internal energy $ \rho T /(\gamma -1)$ and the kinetic 
energy $\rho v^2/2$.
 The terms in the brackets on the left hand side of equation (\ref{ns3}) are the various forms
of energy flux in to or out of a unit volume fluid parcel. 
The first three terms constitute  $(E + p)\bf{v}$, which equals 
the convective flux $c_p \rho T {\bf v}$ plus the
kinetic energy flux $\rho v^2{\bf v}/2$. Away from the upper and
lower boundaries $(E + p)\bf{v}$ represents the energy transported by the resolved large scale eddies.   
The LES model is designed so that this term  carries   most of the vertical energy flux.
The other  fluxes from left to right, are  the viscous flux and  the diffusive flux, $\bf f$. 
The last term in the energy equation is the work done by buoyancy.

At the base ${\bf f}$ has a positive  constant value.  
This is the source term of the vigorous turbulent 
convection  i.e at $r=r_{\rm b}=0.72$, ${\bf f } = f_{\rm b} {\hat{\bf r}}$.
At all other horizontal levels ${\bf f}$ acts as a diffusion term, computed as 
${\bf f } = -k_1 \na S -k_2 \na T$. The
values of $k_1$ and $k_2$  determine whether the layer is convective (unstable) or radiative (stable). In the 
unstable layer ($0.72 \leq r <  0.986$), $k_1=\mu T /{\rm Pr}$
and $k_2 \ll f_{\rm b} /|\na T|_{t=0}$.  
As $k_2$  is very small, nearly all of the heat transport is convective. In the
upper layer ($0.986 \leq  r \leq  1.0$), $k_1=0$ and $k_2 = f_{\rm b} /|\na T |_{t=0}$, allowing conduction to transport
all of the heat flux in the stable layer. The conduction layer emulates radiation above
the convection layer. 

  The horizontal boundaries are insulating, stress free and impenetrable in latitude, and
periodic in longitude. The top and bottom are both impenetrable and stress free. 
The source term  $f_{\rm b}$  is injected in at the bottom and conducted out through the top.
This requires the top to be maintained at a constant temperature.

\subsection{LES versus DNS}
In the SCZ, as the convective flux $\gg$  diffusive flux, efficient mixing reduces the 
superadiabatic gradient $\na -\na_{\rm ad}$ to just above zero ($\approx 10^{-8}$).  
This type of convection
is numerically simulated,  by transferring energy from the base to the top of the computational domain
as follows. Firstly, in the unstable layer (from  $r=0.72$  up to  $r=0.986$)
the gas conductivity is artificially small, forcing convection to carry most of the heat flux 
across the imposed temperature gradient. Secondly, the initial polytrope is neutrally stable
($\gamma  = 5/3, n = 1.5$). After the convection has developed, the relaxed thermal structure
remains very close to this initial state, so that $\na$ is only slightly greater than $\na_{\rm ad}$ (or equivalently   
$\na S$ is just below zero).
  This approach to modeling the SCZ is very different from the DNS approach. 
The main difference lies in the role of the diffusive flux.
In the DNS by Brummel et al. (1996) , radiation transports    
75$\%$ of the energy flux,
while  the resolved  convective motions carry the rest (via the  
kinetic and enthalpy fluxes). In  the present LES, the thermal conductivity is  nearly zero,   
therefore even the entropy diffusion of the SGS, is greater than radiative diffusion.
As $\na S \approx 0$, the resolved large eddies 
must carry the majority of the energy flux.

\subsection{Angular momentum conservation}

Due to the non-zero reference rotation rate, after relaxation one must ensure that there is no mean motion
between the fluid in the shell and the rotating frame of reference. After the start of the computation,
the compressible bulk can acquire some spurious form of rotation due to initial
expansion or contraction. Enforcing the condition ${\langle \rho v_\phi \rangle }_{\rm v} =$ 0, 
where `v' denotes volume averaging
taken  after thermal relaxation, ensures the total angular momentum is zero in the 
reference frame. This is accomplished
by calculating the mean angular velocity of the shell, 
${\langle \Omega  \rangle }_{\rm v} = \Sigma \rho v_\phi r {\rm sin} \theta / \Sigma \rho r^2 {\rm sin}^2 \theta$, 
and subtracting the residual angular momentum from the  
total flow,  $\rho v_\phi \rightarrow   \rho v_\phi - {\langle \Omega  \rangle }_{\rm v} \rho r^2 {\rm sin}^2 \theta$. 
When the statistics are gathered,${\langle \Omega  \rangle }_{\rm v}$ is  less than $4 \times  10^{-4}$.

\subsection{Numerical methods}

After some transformations to make the numerical scheme conservative and preserve second
order accuracy for the nonuniform vertical grid (Chan \&  Sofia 1986), equations (\ref{ns1}-\ref{ns4}) are discretised in
spherical coordinates. Using a code developed by Chan and Wolff (1982), an implicit scheme (the Alternating
Direction Implicit Method on a Staggered grid or ADISM) relaxes the fluid to a self consistent
thermal equilibrium.  The relaxation time is of the order of a `Kelvin Helmholtz' time ($\approx \int \rho c_v T  dr / f_{\rm b}$). 
In the  fully relaxed layer, the energy flux leaving the 
top of the shell is within 5 $\%$ of the input flux $f_{\rm b}$. 

Next a second order explicit method (Adams Bashforth time
integration) gathers the statistics of the time averaged state. The statistical integration
time is over 500 turn over times, and requires about 1.5 million time steps. For model A
(longitudinal span of $30^\circ$), there are  $39\times 70 \times 35$ grid points, in the radial , latitudinal
and zonal directions respectively. For model B which has twice the longitudinal span of A,
there are twice as many points in the longitudinal direction. The choice of grid 
comes from comparing a simulation  in a very small section of the  shell (Robinson 1999) ($\theta \pm 15^\circ$), with 
a physically similar LES computation in a small box (Chan 1999). The box, which has an aspect ratio of 1.5,
is placed at the mid-latitude of the shell.

For a single processor on the ORIGIN2000, the CPU time
per integration step is about 3 seconds in model B. Using automatic parallelization on 4 processors, the
speed up factor is about 3. Numerical stability requires a non-uniform grid with the same
number of grid points per scale height  and
a very small input flux $f_{\rm b}$ of 0.25/64. These  significantly increase the total (implicit plus explicit) 
computation time. 
Consequently, the minimum time for a full simulation is about
a month.

\section {Results}
\subsection{Trials}
\label{trials}
After a series of  numerical experiments (Robinson 1999) in a shell
spanning $\pm 15^{ \circ}$ in
latitude and longitude,
a set of parameters 
were found that  generated  a `sun-like' rotation  pattern. These were  $\Omega_{\rm o} =2.91,
 f_{\rm b}= 0.25/64, {\rm Pr}=1$ and $ g_{\rm t} = 19$ (about 5 pressure scale heights).
Fixing these parameters, 
the latitudinal span
was increased to $60^\circ$ ($30^\circ$ above and below the equator).
Surprisingly,  the rotation profile appeared to
be in some kind of  `quasi-steady state'. Initially a `sun-like'  profile was seen, but
as the computation progressed, the radial angular velocity gradient at the equator,  switched from positive to negative.
This simulation was run to completion and  
the averaged results are classified as model A. 

Increasing $\Omega_{\rm o}$ did not improve matters, and only when the longitudinal
span was increased to $60^{ \circ}$, was a positive angular velocity gradient sustained. 
The shell now covered the same longitudinal as latitudinal extent. 
This simulation was run independently of model 
A and 
the 
averaged results are classified as model B.
We found out later that the flow reversal in A, was caused by a spurious meridional circulation.
Strong downflows  at the impenetrable latitudinal boundaries 
generate a powerful flow, pointed from the equator towards the poles.
This feature is described in section 
\ref{lat}.

\subsection{Flow characteristics}
In
a turbulent fluid a quantity $q$ can be split into a mean and a fluctuating part,
\be
q = \overline{q}(r,\theta)+ q'(r,\theta,\phi,t).
\ee
The overbar represents a  combined longitudinal and temporal average, i.e.
\be
\overline{q}(r,\theta) = \frac{1}{t_2-t_1} \int\limits_{t_1}^{t_2} \left(
 \frac{1}{
 (\phi_2-\phi_1)
}
\int q   d \phi  \right)dt.
\ee
$t_1$, is a time  after the system has reached a self-consistent
thermal equilibrium i.e.  $t_1 >  \int e dr /f_{\rm b}$. $e$ is the internal energy at each horizontal level. 
The time required for statistical convergence ($t_2-t_1$) depends on the
particular quantity being averaged. While the mean velocities required about 100 turnover times, 
turbulent quantities such as the velocity correlation $\overline{ v_r' v_\phi'}$,
took about 5 times as long.

\subsubsection{Mean zonal flow}

The angular velocity averaged over time and longitude, relative to the rotating frame of reference,
is computed and shown in Figs \ref{contphi30} and \ref{avyphi30} for model A, and in Figs
\ref{contphi60} and \ref{avyphi60} for model B. In model A, the isorotation contours are parallel
to the rotation axis. This means the angular velocity of a fluid element at any point in
the shell is determined by its perpendicular distance from the rotation axis. Fig. \ref{avyphi30} shows the angular velocity 
plotted against nondimensional depth (given as a fraction  of the total solar  
radius, i.e. $r = $ r/R)  
 in the Northern hemisphere. Co-latitudes
($90^\circ$-latitude) of $90^\circ, 85^\circ, 79^\circ$ and $67.5^\circ$ are labelled by solid, long-dashed, triple-dot dashed
and dot-dashed curves, respectively. The plots show that the angular velocity decreases
radially outwards and away from the equator, in direct contrast to the rotational profile
found in the SCZ.  

If the simulation is repeated with twice the longitudinal span, the differential rotation has 
a much more `sun-like' appearance. 
In model B, the shape of the isorotation contours resemble helioseismology observations in two
distinctive ways. 

Firstly, an initial increase then decrease in angular velocity from the top
inwards near the equator is found, as shown by the two closed circular rotation contours (or equivalently, the 
radial angular velocity profile in Fig. \ref{avyphi60}).
Helioseismic results, Kosovichev et al (1997), show that the
increase inwards of angular velocity occurs at the equator and
up to 60 degrees co-latitude, but the
jury is still out on the behavior at higher co-latitudes, 
Schou et al (1999). As the computational shell only extends to 
about 60 degrees co-latitude,
we will only consider behavior away from the boundaries  as being representative of the 
actual convective flow. In that sense, 
within $\pm 8^\circ$ about the equator,  the computed angular velocity `bump'                                           
is at least qualitatively similar to the                                  
observed result.

Secondly, away from the equator towards mid-latitudes, the contours are about half way
between the cylindrical contours (Taylor Columns) seen in most earlier global simulations
(e.g. Glatzmaier 1987), and the cone-like shape observed in the SCZ. There is also good agreement with 
recent anelastic global LES by Elliott, Miesch \& Toomre (2000). Over the range of depth and latitude in common with
their simulation and the present simulation, the isocontours are similar. The amount of
variation of angular velocity with latitude is also in agreement with the SCZ. At the same
colatitudes as in A, the mean angular velocity is plotted against depth, Fig. \ref{avyphi60}. At the top
of the shell the angular velocity, drops by about 0.2 between the equator (solid line) and co-latitude of
$67.5^\circ$ (dot-dashed line). As $\Omega_{\rm o}$, is about 3, the drop implies a $7\%$ variation in rotation rate
over $22.5^\circ$, or extrapolating, a pole that spins $28\%$ faster than the equator. 
This is in rough agreement with the 
rotation rate at the surface of the Sun, which 
varies from 25 days at the equator to 35 days at the poles.

\subsubsection{Mean meridional flow}
\label{lat}
The meridional flows are also very different. In A, a strong meridional circulation develops,
directed from the equator towards the poles at the top, with strong downflows at the
latitudinal boundaries. The mean meridional velocity $\overline{v_\theta}$ at the same colatitudes as used for
the zonal flow, is presented in Fig. \ref{avxA}.
As $\overline{v_\theta} < 0$ in the Northern hemisphere, the upper leg of the circulation is directed  from the
equator to the pole. Moving polewards from the equator, where by
symmetry the meridional velocity is zero, $|\overline{v_\theta}|$ increases to a maximum near a colatitude of
$67.5^\circ$, and then reduces to zero as the flow approaches the latitudinal boundary.  At the boundary, the magnitude of the maximum
downward  radial velocity (not shown) 
is about 0.2.  
The meridional circulation is produced by strong downflows associated with
the stress free impenetrable boundaries. In early models of turbulent compressible  convection in a small box,
Chan \& Sofia (1986), noticed that an impenetrable side boundary tends to attract downflows
and they made all side boundaries periodic in later computations. In a shell which excludes
the poles,
impenetrable 
boundaries are unavoidable. These strong downflows create an artificially large meridional circulation.

Fortunately the downflows can be  almost eliminated by widening the longitudinal span. While  
the magnitude of the maximum radial velocity at the boundary  is 0.2 in A, it is less than 0.005
in B (both being directed downwards). The impenetrable boundaries have very little effect
on the downflows in B. This is because less restriction is placed on the flow direction. In
A,  the narrowness of the slice enhances the meridional circulation, enabling it to  traverse the entire depth 
of the shell. 
In B, the meridional flow is only significant in the upper part of the 
shell ($r  >  0.95$), elsewhere it is close to zero. 
Fig. \ref{avxB} shows that $\overline{v_\theta}$ is negative (poleward)  at the top of the unstable layer, while just 
above, in the stable layer, there is a returning (equatorward) flow. The 
return flow is twice as fast as the poleward flow, because the fluid at the top  
has half the density of the equatorward moving fluid, i.e. momentum is conserved in the meridional velocity loop.
The point where $\overline{v_\theta}$ changes 
sign ($r = 0.975$) is very close to  the beginning of the stable layer ($r = 0.985$).  
Below a depth of about about 0.95, $\overline{v_\theta}$ is very small and the multi-cellular
appearance is most likely a residual effect.  

\begin{figure}
\vspace{5.5cm}
\caption{\label{contphi30} Isorotation contours averaged over time and longitude in a shell
spanning $60^\circ$ in latitude and $30^\circ$ in longitude (model A).}
\vspace{5.5cm}
\caption{\label{avyphi30} Depth variation of mean angular velocity
in a shell
spanning $60^\circ$ in latitude and $30^\circ$ in longitude (model A). Co-latitudes of $90^\circ, 85^\circ,
79^\circ $ and $ 67.5^\circ$ are denoted by solid, dashed, triple-dot-dashed and
dot-dashed lines respectively.}
\end{figure}

\begin{figure}
\vspace{5.5cm}
\caption{\label{contphi60} Isorotation contours averaged over time and longitude in a shell
spanning  $60^\circ$ in latitude and longitude (model B).}
\vspace{5.5cm}
\caption{\label{avyphi60} Depth variation of mean angular velocity
in a shell
spanning  $60^\circ$ in latitude and longitude (model B). Co-latitudes of $90^\circ, 85^\circ,
79^\circ $ and $ 67.5^\circ$ are denoted by solid, dashed, triple-dot-dashed and
dot-dashed lines respectively.}
\end{figure}

\begin{figure}
\vspace{5.5cm}
\caption{\label{avxA} Depth variation of mean meridional  velocity
in a shell
with a $30^\circ$ longitudinal span (model A). Co-latitudes of $90^\circ, 85^\circ,
79^\circ $ and $ 67.5^\circ$ are denoted by solid, dashed, triple-dot-dashed and
dot-dashed lines respectively.}
\vspace{5.5cm}
\caption{\label{avxB} Depth variation of mean meridional  velocity
in a shell
with a $60^\circ$ longitudinal span (model B). Co-latitudes of $90^\circ, 85^\circ,
79^\circ $ and $ 67.5^\circ$ are denoted by solid, dashed, triple-dot-dashed and
dot-dashed lines respectively.}
\end{figure}

\subsubsection{Turbulent quantities}

The  root-mean-square (r.m.s.) variance of a quantity $q$ is given by   
\be
q''= \sqrt{\overline {q^2}-{ \overline {q}}^2}.
\ee
We can remove the radial dependence by averaging over the convection 
layer, 
\be
\langle q'' \rangle = \frac{1}{d}\int q'' dr,
\ee
where $d=1 - r_{\rm b}$ is the total depth of the layer. 
The angled brackets will be used to denote depth averages.  
Note $\langle q'' \rangle$ depends only on latitude.

Turbulent velocity characteristics at co-latitudes of $90^\circ , 84^\circ$ and $67.5^\circ$ 
are presented in tables \ref{turbA} and \ref{turbB} for the simulations  A and  B.   
The total kinetic energy (mean plus turbulent)  is proportional to  
${\overline{ \langle v^2 \rangle} } ^ {1/2}$, while the  
turbulent part of the kinetic energy is measured  by 
$\langle v'' \rangle$, where $v''=\sqrt{{v_r''}^2+{v_\theta''}^2 + {v_\phi''}^2}$.
Columns 2 and 3 indicate the relative sizes of 
${\overline{ \langle v^2 \rangle} } ^ {1/2}$ and $\langle v'' \rangle$. 
In A, ${\overline{ \langle v^2 \rangle} } ^ {1/2}$ is more than 4 times greater than 
$\langle v'' \rangle$, i.e. most  of the kinetic energy is in the mean circulation. 
Conversely in B,  $\langle v'' \rangle$ is within a factor of 2 
of ${\overline{ \langle v^2 \rangle} } ^ {1/2}$, implying the kinetic energy is split about equally between the 
mean and turbulent scales. 
In model B, most of $\langle v'' \rangle$ is in the zonal 
component $\langle v_\phi'' \rangle$, 
while in A,  
$\langle v_\phi'' \rangle$ is the smallest component.  
The distribution of kinetic  energy  between the mean and turbulent flow may determine                        
the nature of the differential rotation. 
By measuring the autocorrelation function of the vertical velocity, 
we found that the vertical scale of the turbulence is about  1.5 pressure 
scale heights (P.S.H), compared to a total shell depth of 5 P.S.H.

Nondimensional parameters defining the importance of rotation compared to other
physical processes can be calculated from the
resultant flow.  The strength of the turbulent convection
relative to rotation is characterized by the Coriolis number, ${\rm Co} =\Omega_{\rm o} d / \langle v'' \rangle $.
The  Reynolds number, ${\rm Re}$, compares the relative magnitudes of the
advection and  viscous terms, and is calculated as $ \langle v'' \rangle d /\langle \overline {\mu}/\overline {\rho}\rangle $.
The importance of SGS viscosity relative to rotation, is measured by the
Taylor number, ${\rm Ta}$, which is $4\Omega_{\rm o}^2 d^4 / {\langle \overline {\mu}/\overline {\rho}\rangle}^2$
For reference these non-dimensional parameters are computed from  the relaxed flow 
and presented in the last three columns of the tables \ref{turbA} and \ref{turbB}.
The Coriolis numbers are larger in A because the turbulent velocities 
are smaller.  This suggests rotation will have a greater  effect on the 
large scales in A than in B. The larger Reynolds numbers in B reflects the more turbulent nature of  the flow.
\begin{table*}
\begin{minipage}{140mm}
\caption{\label{turbA}Model A: Turbulence characteristics in the `non sun-like' differential rotation.}
\begin{tabular}{ccccccccc}
$ \theta^\circ$ & $ {\langle  \overline {v^2} \rangle}^{1/2}  $ & $  \langle v'' \rangle$ &$\langle {v_r}'' \rangle$&
$ \langle {v_\theta}'' \rangle $ &$ \langle {v_\phi}'' \rangle$ &
Co & Re & Ta\\
 & & & & & & & & \\
67.5&  0.361 & 0.094 & 0.064 & 0.059 & 0.035  & 8.6 & 330 & 2.1e+08 \\

84  & 0.569 & 0.076 & 0.056 & 0.043 & 0.024  & 10. & 290 & 2.5e+08 \\

90  & 0.591 & 0.079 & 0.059 & 0.042 & 0.028  & 9.8 & 270 & 2.0e+08 \\
\end{tabular}
\end{minipage}
\end{table*}

\begin{table*}
\begin{minipage}{140mm}
\caption{\label{turbB}Model B: Turbulence characteristics in the `sun-like' differential rotation.}
\begin{tabular}{cccccccccc}
$ \theta^\circ$ & $ {\langle  \overline {v^2} \rangle}^{1/2}  $ & $  \langle v'' \rangle$ &$\langle {v_r}'' \rangle$&
$ \langle {v_\theta}'' \rangle $ &$ \langle {v_\phi}'' \rangle$ &
Co & Re & Ta\\
  & & & & & & &  \\
67.5&  0.157 & 0.105 & 0.056 & 0.053 & 0.070  & 7.4 & 440  & 2.9e+08 \\

84  & 0.238 & 0.122 & 0.065 & 0.063 & 0.080  & 6.3 & 480  & 2.6e+08 \\

90  & 0.250 & 0.121 & 0.066 & 0.060 & 0.079  & 6.4 & 460  & 2.5e+08 \\
\end{tabular}
\end{minipage}
\end{table*}
\section{Discussion: What produces the differential rotation?}
\subsection{Taylor-Proudman balance (TPB)}
Averaging the meridional momentum equation 
over longitude and time, 
so that $\pa / \pa \phi=0$ and  $\pa / \pa t=0$, produces 
\\
\bea
\frac {1}{r^2}\frac{\pa}{\pa r }( \overline{\rho v_\theta v_r}r^2)
&+&\frac {1}{r {\rm \sin} \theta} \frac{\pa}{\pa \theta } ({\rm \sin} \theta
\overline{\rho v_\theta^2 }) 
\nonumber
\\
&+& \frac {1}{r}
(\overline{\rho v_\theta v_r} - {\rm \cot} \theta
\overline{\rho v_\phi^2}) 
\nonumber
\\ 
& +& \frac {1}{r} \frac {\pa \overline{P}}{\pa \theta}  -
2 \Omega \overline{\rho v_\phi}{\rm \cos} \theta
+ V_\theta = 0
\label{sheq1}
\eea
\\
Measured at an arbitrary  co-latitude  of $78^\circ$, Figs \ref{phi30i21} and \ref{phi60i21} 
show the relative sizes of terms in (\ref{sheq1}) computed from the averaged flows in A and B. From left to right
terms  are labeled by
pluses, stars, diamonds, triangles, boxes and crosses, respectively. The third 
term in each equation is split into
two terms (denoted by the diamonds and triangles). For the Co,  Re and Ta values 
in  the  present simulations,  
the  viscous stress terms 
$V_\theta$ are, the equator excluded,  much smaller than other
terms.  Therefore these terms  are not shown. 

In both A and B the   
latitudinal pressure gradient (boxes)  and the Coriolis
force (crosses) are in approximate balance. The terms 
which contain the Reynolds stresses are 
insignificant. 
The overall dominance of the pressure gradient and the Coriolis force suggest the 
Navier Stokes momentum equations can be approximated by: 
\be
\frac{\na P }{\rho} = {\bf g} -{\rm 2} {\bf \Omega_o}  \times { \bf v}.
\label{momtum}
\ee
The above equation is known as the Taylor-Proudman momentum balance. Using entropy
$S = c_v \ln(p/\rho^\gamma)$ and the perfect gas equation $p=\rho {\rm R} T$, the curl of the 
left hand side of (\ref{momtum}) 
produces:
\be
\na \times \frac{\na P}{\rho}=-\frac{1}{\rho ^2}(\na P \times \na \rho)= \na T 
\times \na S,
\label{TS}
\ee
so that the $\phi$ component of the curl of equation (\ref{momtum}) is equivalent  to
\be
2 \Omega_o \biggl(r {\rm \cos} \theta \frac{\pa v_\phi}{\pa r}
- {\rm \sin} \theta \frac{\pa v_\phi}{\pa \theta}\biggr)  + \frac{\pa T}{\pa r}\frac{\pa S}{\pa \theta} - 
\frac{\pa T}{\pa \theta} \frac{\pa S}{\pa r}=0.
\label{eqdurney}
\ee 
As ${\bf g}$ can be written as a potential, the buoyancy term disappears.

Measurements from the simulation show that  ${\pa T}/{\pa r}$ is nearly three 
orders of magnitude greater than ${\pa T}/{\pa \theta}$.
The temperature field is almost radial 
and does not alter much from the initial unperturbed state.
Assuming $T \approx T(r)$,   
the last term of (\ref{eqdurney}) can be neglected, and the second term can be replaced by
\be
\frac{dT}{dr} \frac{\pa S}{\pa \theta}=-\frac{g(r)}{c_{\rm p}} \frac{\pa S}{\pa \theta}
\ee
which when substituted into (\ref{eqdurney}), gives 
\be
 2 \Omega_o \biggl(r {\rm \cos} \theta \frac{\pa v_\phi}{\pa r}
- {\rm \sin} \theta \frac{\pa v_\phi}{\pa \theta}\biggr) - \frac{g(r)}{c_{\rm p}} \frac{\pa S}{\pa \theta}=0. 
\label{TPB}
\ee
This is equivalent to the rotation law obtained by Durney (1999).

Fig. \ref{smn} shows the mean entropy, $\langle \overline{S} \rangle$ averaged over time, longitude and depth,  plotted
against co-latitude for models A (triple dot dashed line) and B (solid line).
In A, the entropy is almost independent of co-latitude, suggesting that 
$\pa S / \pa \theta \approx 0 $, so that equation (\ref {TPB}) reduces to
\be 
r {\rm \cos} \theta \frac{\pa v_\phi}{\pa r}
- {\rm \sin} \theta \frac{\pa v_\phi}{\pa \theta} =0.
\label{tp} 
\ee
Equation (\ref{tp}) implies $v_\phi=f(r {\rm sin}\theta$), i.e. $v_\phi$  is constant along cylinders parallel
to the rotation vector. In A  because the pressure gradient and Coriolis terms dominate the 
momentum balance, and the latitudinal entropy gradient is approximately zero, Taylor columns
are seen in the interior. This explains the isorotation contours in Fig. \ref{avyphi30}.
  
Contrastingly, in B the entropy varies significantly with colatitude.
This non-zero latitudinal entropy gradient is the reason  why  more `sun-like'
isorotation contours are seen in model B. In other words, the term $-g(r)/c_{\rm p} \cdot \pa S / \pa \theta $ shapes the
differential rotation. Following this line of reasoning, we computed the baroclinic term $-g(r)/c_{\rm p}  \cdot \overline {\pa S / \pa \theta}$
(denoted
by boxes) and  $2\Omega_{\rm{o}}(r {\rm \cos} \theta \overline{\partial v_\phi/\partial r}
- {\rm \sin} \theta \overline{\partial v_\phi/\partial \theta})$ (denoted by crosses) from the averaged flow.
Fig. \ref{tpbterms} shows the results at co-latitudes of $90^\circ$ (smallest magnitude), $82^\circ$ and $72^\circ$ (largest
magnitude). Clearly they are in approximate balance, confirming that the
{\it thermal structure i.e.  $S(r,\theta)$, directly controls the overall shape of the differential
rotation profile.  The Reynolds stresses  play a only  minor role in the dynamics}. 
As the simulation is of `mildly', rather than fully developed turbulence, it remains to be seen, whether the same is true in the sun. 
\begin{figure}
\vspace{5.5cm}
\caption{\label{phi30i21} Successive terms in averaged meridional momentum equation
measured at a co-latitude of $78^\circ$ in a shell
with a $30^\circ$ longitudinal span (model A).
Clearly the  largest terms are the latitudinal pressure gradient (boxes) and the
Coriolis force (crosses), while the Reynolds stresses are relatively insignificant.}
\vspace{5.5cm}
\caption{\label{phi60i21} Successive terms in averaged meridional momentum equation
measured at a co-latitude of $78^\circ$ in a shell
with a $60^\circ$ longitudinal span (model B).
Similarly  the  largest terms are the latitudinal pressure gradient (boxes) and the
Coriolis force (crosses) and again the Reynolds stresses are very  small.}
\end{figure}
\begin{figure}
\vspace{5.5cm}
\caption{\label{smn}Longitude, time and depth averaged entropy variation with co-latitude: simulations A and B are denoted by
triple dot dashed and solid lines,  respectively. The entropy is
almost constant
in model A, but has a significant latitudinal variation
in B. In both cases the entropy gradient is zero at the equator.}
\vspace{5.5cm}
\caption{\label{tpbterms} Averaged terms in the Taylor-Proudman balance for the shell with a $60^\circ$ longitudinal span (model B).
 $-g(r)/c_{\rm p} \cdot
\overline {\partial S/\partial \theta}$ and $2\Omega_{\mathrm{o}}(r {\rm \cos} \theta \overline{\partial v_\phi/\partial r}
-{\rm \sin} \theta \overline{\partial v_\phi/\partial \theta})$,  are denoted by boxes and crosses respectively.
Both  expressions   have the smallest magnitude at the equator and successively larger values
at co-latitudes of $82^\circ$ and $72^\circ$.}
\end{figure}

\subsection{Meridional circulation and the Reynolds stress}
\subsubsection{Meridional circulation produces differential rotation in A}

In model A, meridional circulation drives the convection to 
a spurious  equilibrium state (see sections \ref{trials} and \ref{lat}). The computation starts off with a sun-like 
rotational state,  but as it progresses 
the layer relaxes to a completely different equilibrium.
We previously suggested that this is  a consequence  of the artificially 
large downflows at the impenetrable latitudinal boundaries.
We will now show that the direction  and strength of the meridional circulation in A is 
just enough  to drive the differential rotation. 

Consider a fluid parcel in a Lagrangian frame of reference. Ignoring 
the zonal pressure gradient and frictional effects, the only 
force on the parcel is the Coriolis force associated with the 
meridional circulation $(v_r,
 v_\theta, 0)$. The equation of motion of the fluid parcel is then
\be
d v_\phi/dt \approx -2\Omega_{\rm o}{\rm cos}\theta v_\theta  -2\Omega_{\rm o}{\rm sin}\theta v_r.
\ee
Integrating w.r.t time gives
\be
\Delta v_\phi \approx -2\Omega_{\rm o} \Delta ({\rm sin}\theta)   -2\Omega_{\rm o}{\rm sin} \theta  \Delta r  
\label{cons}
\ee
We can verify  this relation by taking values of $\overline{v_{\phi}}$ from Fig.
\ref{phi30zonal} which shows how  $\overline{v_{\phi}}$ varies with depth in model A.
At the top of shell (constant $r$),   $\Delta v_\phi/\Delta ({\rm sin}\theta) \approx  5.9 $.
While at the equator (constant $\theta$), 
$\Delta v_\phi / {\rm sin} \theta \Delta r \approx -5 $. As $\Omega_{\rm o}=2.91$,  the meridional circulation is 
about the right size and direction to produce the zonal velocity variation in A.  
This expresses as a tendency of fluid parcels to conserve their individual angular momentum.

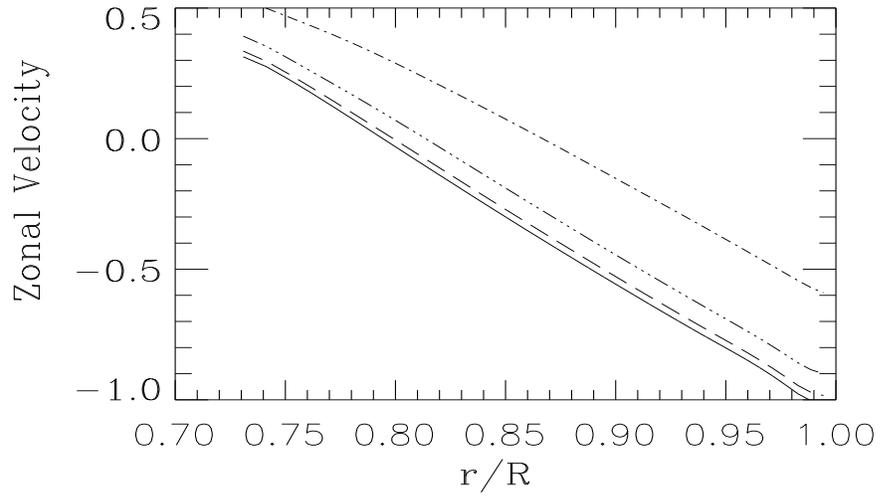
\begin{figure}
\vspace{5.5cm}
\caption{\label{phi30zonal} Longitude and time averaged zonal
velocity in the narrower shell (model A). Co-latitudes of $90^\circ, 85^\circ,
79^\circ $ and $ 67.5^\circ$ are denoted by solid, dashed, triple-dot-dashed and dot-dashed lines respectively.}
\end{figure}

\subsubsection{Reynolds stress produces meridional circulation in B}

In model B, neither the meridional circulation,  nor the Reynolds stress, 
are  sufficient to  directly drive the differential rotation. As 
previously described, it is the entropy distribution which produces the differential rotation in B.
The meridional circulation in B is completely different to that in A. 
Equation (\ref{cons}) cannot be applied to the radial and latitudinal variation of 
zonal velocity in  B. Fluid parcels do not appear to conserve their individual angular momentum.
In fact they do, it is just that other forces are involved, 
namely the Reynolds stresses.

In a turbulent fluid, the velocity can be split into its mean and fluctuating part. Though
the average of the fluctuation is zero, the average of the product of two
fluctuating quantities is not necessarily zero.
The Reynolds stress is the averaged correlation between small scale velocity
fluctuations, in two component directions, and the density.
To reduce the order of the statistical  moments,
we assume that density can be taken out of the correlation,
\be
\overline {\rho v_i'v_j'} = \overline{\rho} \overline{v_i'v_j'}. 
\ee
This approximation is good  because the density fluctuations 
are small (of the order of the square of the turbulence Mach number). 
This has been confirmed numerically. 
As the mean density only depends on  depth, the nature of the 
Reynolds stress can be  ascertained by looking at the  
velocity correlation, $\overline{v_i'v_j'}$.  
Fig. \ref{phi60fvzvy} shows $\overline{v_r' v_\phi'}$ plotted at the same co-latitudes and
using the same line markers
as previous  velocity plots.
Fig. \ref{phi60fvxvy} shows  $\overline{v_\theta' v_\phi'} $, at depths of 0.98, 0.95, 0.90 and 0.85, denoted by 
solid, dotted, dashed and dot-dashed lines, respectively.  

By comparing the plots of $\overline{v_\theta}$ (Fig. \ref{avxB}) and 
the velocity correlations,  certain features become apparent. Firstly, quantities 
are only significant near the top of the shell (above $r = 0.90$), elsewhere 
they are close to zero. Secondly, at lower co-latitudes
(specifically $79^\circ$ and $67.5^\circ$) the sign and order of magnitude of  $\overline{v_\theta}$
closely matches  negative  the slope of $\overline{v_r' v_\phi'}$. 
Thirdly, towards the equator, $\pa / \pa r (\overline{v_r' v_\phi'})$ decreases, while  
$\pa/\pa \theta (\overline{v_\theta' v_\phi'}) $ becomes steeper (more negative).

The relation between  $\overline{v_\theta}$ and the Reynolds stresses (or velocity correlations),
can be traced to the zonal momentum balance.  In a rotating  
frame of reference, the axisymmetric zonal momentum equation can be written as,
\bea 
\frac{1}{r^2} \frac{\pa}{\pa r} (v_r v_\phi r^2)&+&
\frac{1}{r {\rm sin} \theta} \frac{\pa}{\pa \theta} (v_\theta v_\phi {\rm sin}\theta )+
\nonumber
\\
N&+&2 \Omega_{\rm o}(v_\theta {\rm cos} \theta
+v_r {\rm sin}\theta)  \approx  0,
\eea
where $N$ denotes the additional non-linear terms and the viscous terms are again 
excluded. Near the impenetrable top boundary, $v_r$ is close to zero, so the last term can be 
excluded.  
Clearly the derivatives of 
$\overline{v_r' v_\phi'}$ and $\overline{v_\theta' v_\phi'}$ are capable of driving $\overline{v_\theta}$. 
Away from the equator $\overline{v_r' v_\phi'}$ drives $\overline{v_\theta}$, while close to the equator (specifically    
$85^\circ$) 
$\overline{v_\theta' v_\phi'}$ becomes more important. 
  
Overall, the Reynolds stresses produce a weak meridional circulation 
concentrated in the uppermost part of the shell,
but are negligible elsewhere. 

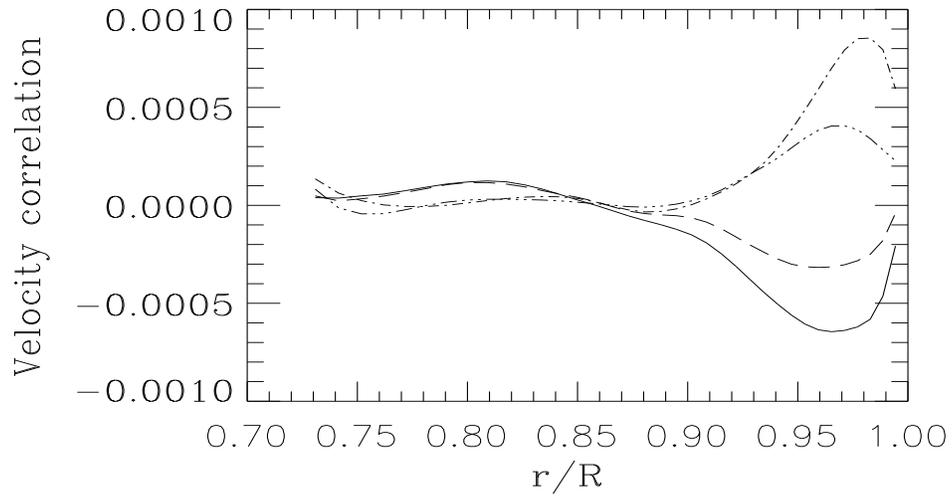
\begin{figure}
\vspace{5.5cm}
\caption{\label{phi60fvzvy} Longitude and time average of the product of radial and zonal velocity fluctuations
$\overline{{v'}_r {v'}_\phi}$ ($\approx$ Reynolds stress divided
by mean density)
versus depth (model B). Co-latitudes of $90^\circ, 85^\circ,
79^\circ $ and $ 67.5^\circ$ are denoted by solid, dashed, triple-dot-dashed and
dot-dashed lines respectively. The negative troughs near the equator (solid and dashed lines)
represents an inward transport of angular momentum by the Reynolds stress.
The positive peaks away from the equator (triple-dot-dashed and
dot-dashed lines) correspond to a change in the sign of the meridional velocity, indicating that 
$\overline{v_\theta}$  
depends on the radial derivative $ \overline{{v'}_r {v'}_\phi}$. }
\end{figure}

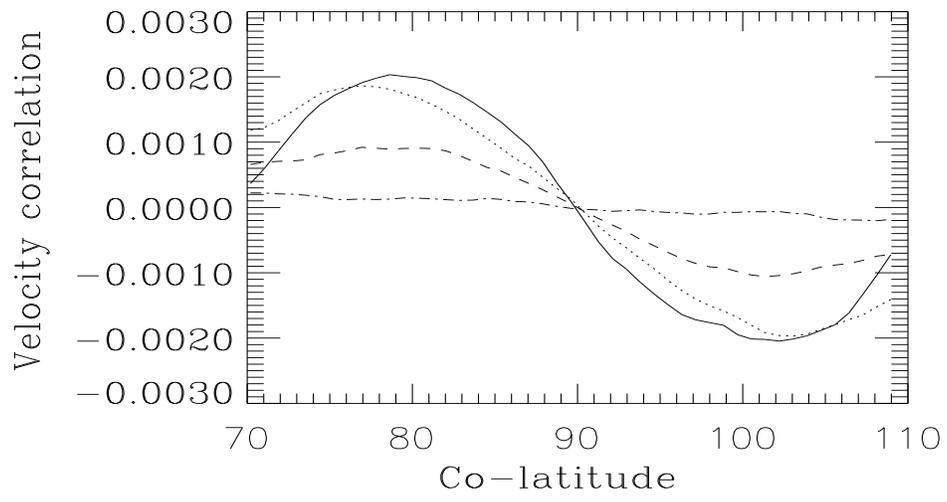
\begin{figure}
\vspace{5.5cm}
\caption{\label{phi60fvxvy} Longitude and time average of the product of meridional and zonal velocity fluctuations
$\overline{{v'}_\theta {v'}_\phi}$ ($\approx$ Reynolds stress divided
by mean density)
versus co-latitude (model B). Depths of 0.98, 0.95, 0.90 and 0.85
are denoted by solid, dotted, dashed and
dot-dashed lines respectively. Near the top (solid and
dotted lines) the
latitudinal gradient of $\overline{{v'}_\theta {v'}_\phi}$ increases towards the equator.}
\end{figure}

\subsubsection{Reynolds stress produces angular velocity `bump' near the top}
\label{rstress}

One other feature of the zonal velocity profile 
that cannot be explained by the  large scale interaction of 
convection with rotation (i.e. TPB), is  
the small angular velocity  `bump' 
(see Fig. \ref{avyphi60}).
The initial  increase in angular velocity moving inwards can be considered as a small 
scale feature of the flow ($\Delta \overline{v_\phi} < 1\%$ of the mean rotation rate). 
The `bump' cannot be attributed to the large scale entropy variation. 

It seems likely that this small scale feature could be caused by the 
Reynolds stress. 
The plot of  $\overline{v_r' v_\phi'}$ (Fig. \ref{phi60fvzvy}),   
reveals  a connection between a negative drop in 
$\overline{v_r' v_\phi'} $ near the top of the shell, 
and the zonal velocity variation (Fig. \ref{avyphi60}).   
The negative trough near the equator  at co-latitudes of 
$90^\circ$ and $85^\circ$ (solid and dashed lines)  represents an inward transport of angular momentum by the 
Reynolds stress. 
We suggest that this  is the cause  of
the slight increase in angular velocity (or `bump') moving inwards from the top of the convection zone.
Furthermore, the maximum angular velocity  at the equator, occurs at almost the same position as the minimum
of $\overline{v_r' v_\phi'}$.

\section{Conclusion}
The aim of this work is to reproduce solar differential rotation, by 
solving the equations of hydrodynamics, in a section of a spherical shell.
To achieve this, we have done two numerical simulations.
The first simulation initially 
has  an angular velocity that  decreases inwards, in agreement with the solar case.
However, as the computation progresses, the
radial angular velocity gradient changes sign, and the rotation rate 
increases inwards.  
The mean rotational structure 
consists of cylindrical
isorotation contours and a sub-rotating equator.
The switch is due to an artificially large
meridional circulation, which itself is a result of strong downflows
that occur at the impenetrable side boundaries. 

In the second computation
the longitudinal span is doubled, so that it now equals the latitudinal span.
The wider shell has much weaker downflows and a more turbulent flow.
Under these  conditions the rotation profile
remains in the `sun-like' state.  The angular velocity
now bears a closer resemblance to solar case, with a radial and latitudinal variation,
both qualitatively and quantitatively similar to the SCZ.

By using an implicit timesteping scheme,  we are able to run the simulations
for longer than a `Kelvin Helmholtz' time scale. 
The emphasis on complete thermal relaxation,
could be the vital ingredient, required to correctly model rotating convection.
Incomplete thermal relaxation, may be one reason why earlier simulations, 
failed to  
get the proper rotation pattern.
In the `Sun-like' simulation,  it is the {\it large}  scale
thermal structure (specifically the entropy variation with latitude),  rather than {\it small} scale motions (as in
mean  field models or some numerical models), that directly produces the
differential rotation. 
The non-zero latitudinal entropy gradient (baroclinicity) 
shapes the differential rotation. 
We notice that in the `sun-like' model, 
the kinetic energy distributes roughly equally between the mean and turbulent scales, 
in contrast to the non `sun-like' model, in which 
nearly all of the energy is in the mean flow. Furthermore, the 
turbulent kinetic energy is greatest in the zonal direction. 
This suggests the turbulent nature of the SCZ, may have some indirect role in the 
maintenance of solar differential rotation, but how operates 
is still unclear.  

The  Reynolds stresses are important in the 
upper 10 \% of the computational domain. This region contains the top of the 
convection (unstable) layer and all of the radiative (stable) layer.
Here the velocity correlation $\overline{v_r' v_\phi'}$, has a dual 
effect. Firstly, it provides an inward transport of angular momentum, which causes the slight
increase in angular velocity, just below the surface and  near the equator. 
Secondly, 
$-\pa / \pa r (\overline{v_r' v_\phi'})$ 
is principally responsible for the poleward meridional circulation at the top of the 
convection layer, and the accompanying return flow in the stable layer. Closer to the equator (within about $5^\circ$),
$-\pa / \pa \theta (\overline{v_\theta' v_\phi'})$ 
appears to drive the meridional flow.
In the meridional momentum equation, the equator excluded,  the only significant terms
are  the pressure gradient and the Coriolois force.
The Reynolds stresses are only important in  the  zonal momentum equation, 
here they generate the meridional circulation found at the top of the shell.

The key result is that the dynamics near the top (surface layers) of the convection zone,  are 
controlled by the Reynolds stresses,
while elsewhere, the differential rotation is determined 
the Taylor-Proudman momentum balance. 
A similar conclusion was reached in the semi-analytical model of the SCZ 
by Durney (2000) (and references therein).

\section*{Acknowledgments}

F.J.Robinson would  like to thank the 
Mathematics Department of the University of Newcastle Upon Tyne 
for their hospitality.
The authors are grateful for comments from 
Bernard Durney which led to significant improvements in this 
work. 
This paper was completed at the Astronomy Department 
of Yale University.

\bsp
\label{lastpage}


\begin{thebibliography}{99}
\bibitem{b1} Becker. E., 1968, Gas Dynamics, Academic Press, p.229

\bibitem{b2} Brummel, N.H., Hurlbert, N.E., Toomre, J. 1996, ApJ 473, 494

\bibitem{b3} Chan, K. L., 1999, ApJ submitted

\bibitem{b4} Chan, K. L. and Sofia, S. 1986, ApJ 307, 222

\bibitem{b5} Chan, K. L. and Wolff, C. L., 1982, J. Comput. Phys. 47,  109

\bibitem{b6} Durney, B.R. 1999, ApJ 511, 945

\bibitem{b7} Durney, B.R. 2000, ApJ 528, 486

\bibitem{b8} Elliott, J.R., Miesch, M.S. and Tommre, J. 2000, ApJ,  533,  546

\bibitem{b9} Gilman, P.A., 1978, Geophys. Astrophys. Fluid Dyn., 11, 157

\bibitem{b10} Glatzmaier, G.A. 1987,
in B.R. Durney and S. Sofia eds.,
The Internal Solar Angular Velocity
Dordrecht: Reidel, p. 263

\bibitem{b11} Kosovichev, A.G., et al  1997  Sol. Phys., 170, 43

\bibitem{b12} Libbrecht, K.G. 1989, ApJ, 336, 1092

\bibitem{b13} Miesch, M.S., Elliott, J.R., Toomre, J., Clune, T.L.,
Glatzmaier, G.A. and Gilman, P.A.  2000, ApJ, 532, 593

\bibitem{b14} Robinson, F. J., 1999, PhD thesis, Hong Kong University of Science and Technology

\bibitem{b15} Scherrer, P.H., et al. 1995, Sol. Phys.,  162, 129

\bibitem{b16} Schou, J., et al. 1999, ApJ, 505, 390

\bibitem{b17} Smagorinsky, J.S., 1963, Mon. Weather. Rev.,  91,  99

\end{thebibliography}
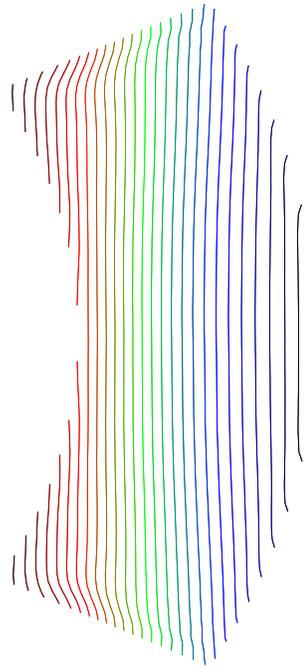
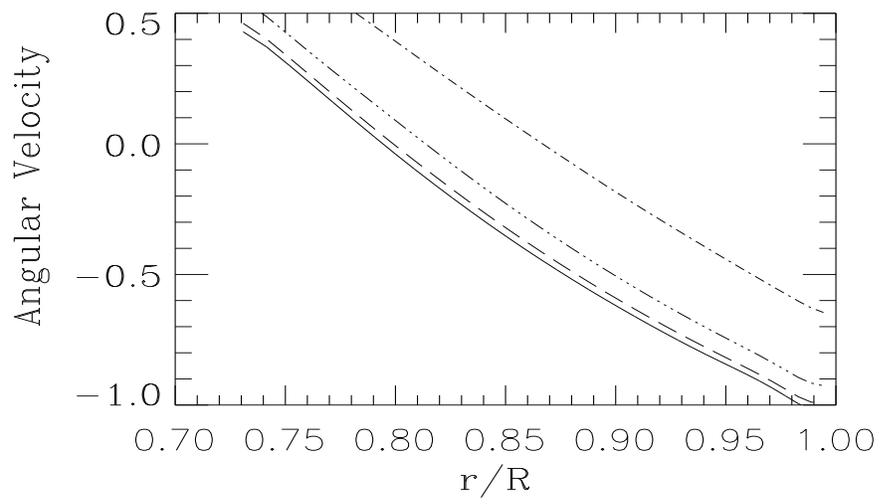
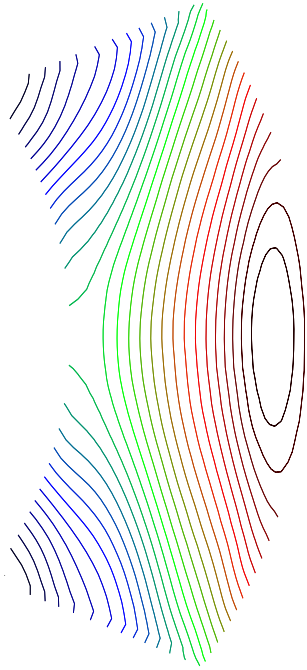
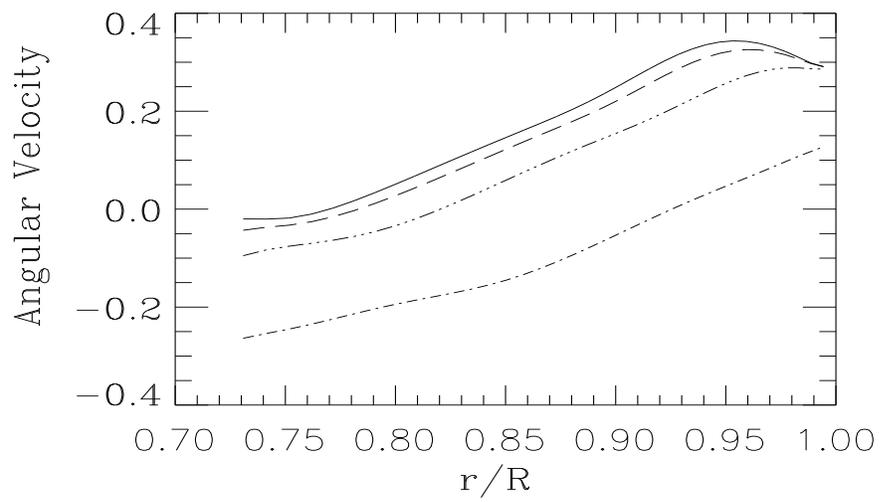
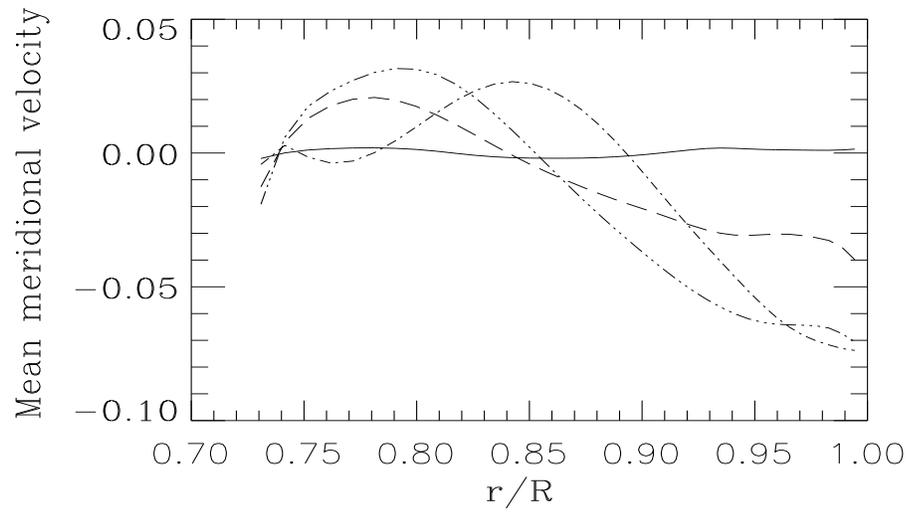
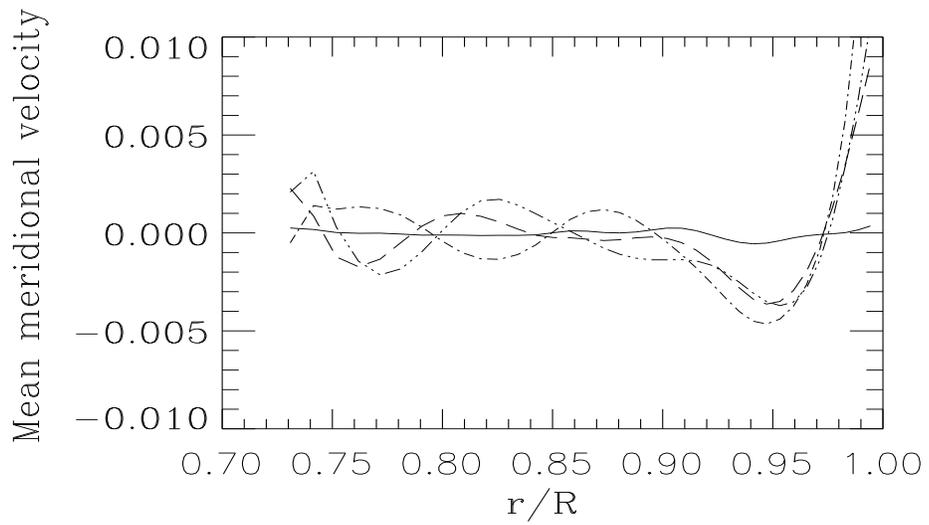
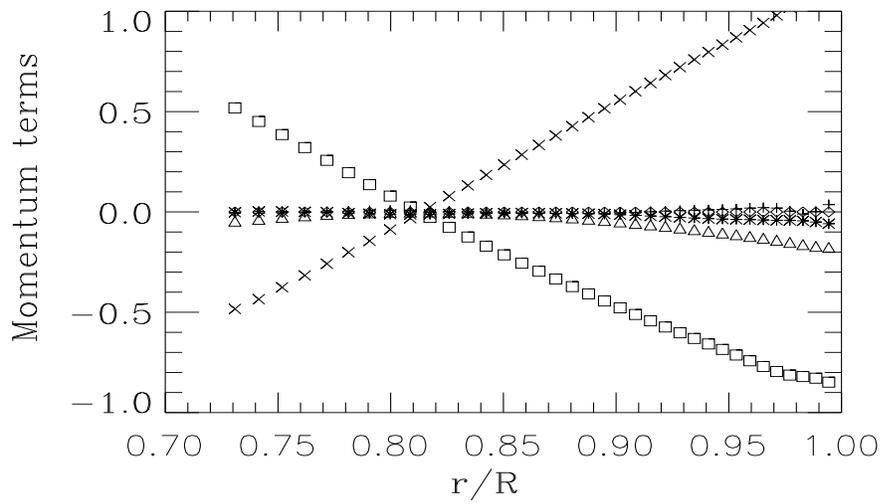
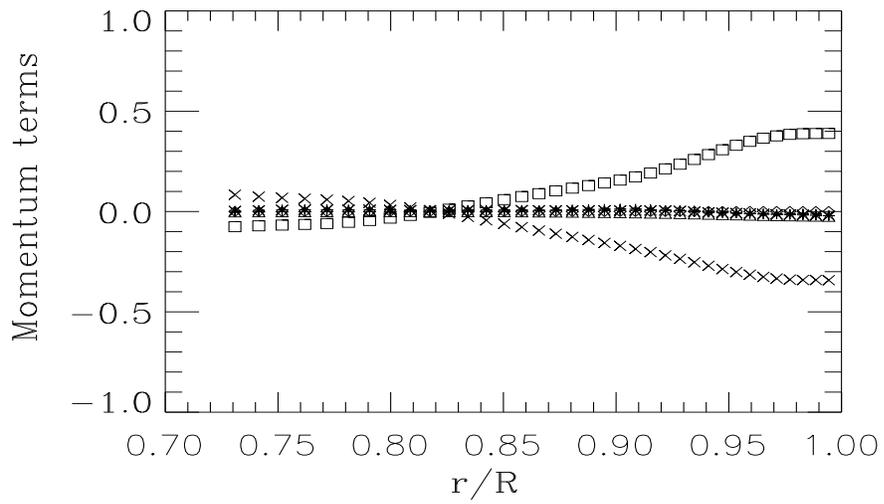
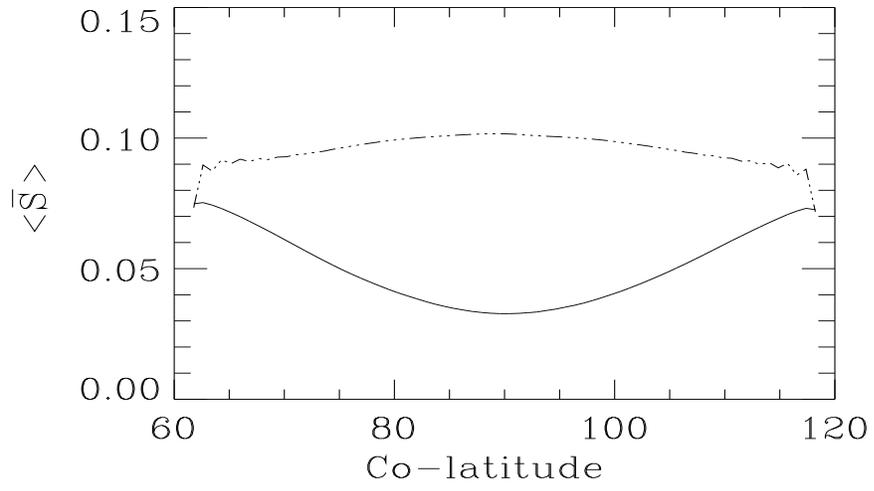
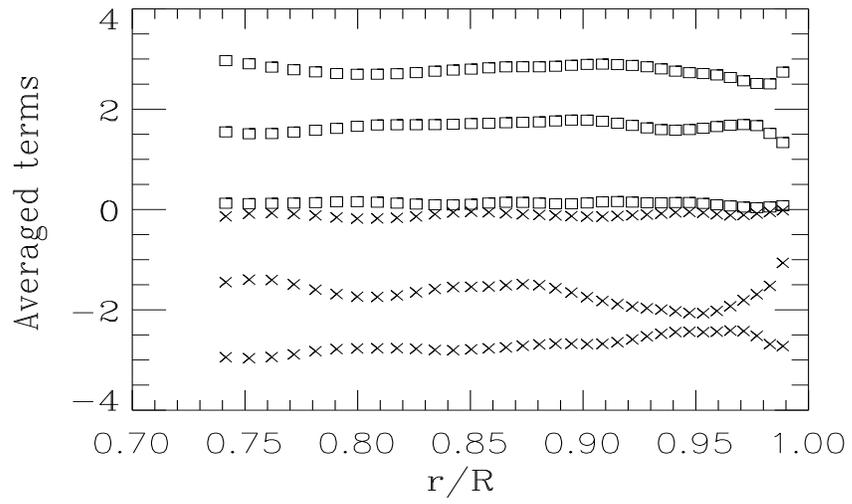
\end{document}